\documentclass[twocolumn,showpacs,preprintnumbers,amsmath,amssymb,superscriptaddress]{revtex4}

\usepackage{graphicx}
\usepackage{dcolumn}
\usepackage{bm}

\begin{document}


\title{Quantum noise in the Josephson charge qubit}

\author{O. Astafiev}
\email{astf@frl.cl.nec.co.jp} \affiliation{The Institute of
Physical and Chemical Research (RIKEN), Wako, Saitama 351-0198,
Japan}
\author{Yu. A. Pashkin}
\altaffiliation[On leave from ]{Lebedev Physical Institute, Moscow
117924, Russia} \affiliation{The Institute of Physical and
Chemical Research (RIKEN), Wako, Saitama 351-0198, Japan}
\author{Y. Nakamura}
\affiliation{The Institute of Physical and Chemical Research
(RIKEN), Wako, Saitama 351-0198, Japan} \affiliation{NEC
Fundamental and Environmental Research Laboratories, Tsukuba,
Ibaraki 305-8501, Japan}
\author{T. Yamamoto}
\affiliation{The Institute of Physical and Chemical Research
(RIKEN), Wako, Saitama 351-0198, Japan} \affiliation{NEC
Fundamental and Environmental Research Laboratories, Tsukuba,
Ibaraki 305-8501, Japan}
\author{J. S. Tsai}
\affiliation{The Institute of Physical and Chemical Research
(RIKEN), Wako, Saitama 351-0198, Japan} \affiliation{NEC
Fundamental and Environmental Research Laboratories, Tsukuba,
Ibaraki 305-8501, Japan}

\date{\today}

\begin{abstract}
We study decoherence of the Josephson charge qubit by measuring
energy relaxation and dephasing with help of the single-shot
readout. We found that the dominant energy relaxation process is a
spontaneous emission induced by quantum noise coupled to the
charge degree of freedom. Spectral density of the noise at high
frequencies is roughly proportional to the qubit excitation
energy.
\end{abstract}

\maketitle

Artificial two-level systems attract interest of researchers
because such systems are believed to be used as quantum bits
(qubits) --- basic elements for quantum computers. Recently much
progress has been achieved in experiments on the Josephson qubits
\cite{Squbit,Saclay,Yu,Martinis1,Delft,Chalmers,Dqubit,CNOT}. All
these experiments have shown that qubits are affected by
decoherence which becomes a key issue of research.

Decoherence of small Josephson circuits has been studied in a
number of theoretical papers (see for example, Ref. \cite{Makhlin,
Averin}). Recently, decoherence of Josephson charge qubits has
been investigated experimentally for some special cases. For
instance, in charge echo experiments, dephasing of the Josephson
charge qubit has been characterized under a bias condition far
from the charge degeneracy point \cite{Dephasing}. Also,
relaxation rate
of charge qubits off the degeneracy point has been measured \cite%
{Schoelkopf,Chalmers}. Dephasing and energy relaxation were
studied in a qubit with the charging-to-Josephson-energy ratio
about unity, for which charge noise may also be important
\cite{Saclay,Decsac}. However, to understand the origin of the
decoherence, a systematic study of the qubits is of great
importance.

In this work, we measure and analyze the qubit energy relaxation
rate in a wide range of parameters. In other words, we use the
qubit as a spectrum analyzer to study the noise of the environment
\cite{Schoelkopf2}. We found that the quantum noise is an
asymmetric charge noise, which causes energy relaxation with
approximately linear frequency dependence ($f$-noise). We also
measure dephasing rate of the qubit and found that it is
consistent with an assumption that it is caused by the 1/$f$
noise.

\begin{figure}
\includegraphics{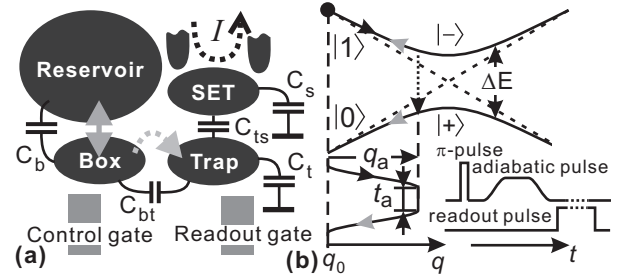}
\caption{\label{fig:T1cFig1} (a) Schematic device representation.
(b) State manipulation diagram for measurements of the qubit
relaxation rate. The inset shows operation pulse sequence.}
\end{figure}

The device schematically shown in Fig.~\ref{fig:T1cFig1}(a)
(identical to that described in Ref.~\cite{RComm}) consists of a
qubit and a readout part. The qubit is a Cooper-pair box
\cite{Squbit} (an island with area of 40 $\times$ 800 nm$^{2}$)
connected to a reservoir through a tunnel junction of SQUID
geometry with Josephson energy $E_{J}$ controlled by external
magnetic field. The readout part contains a charge trap island and
an electrometer
--- a single electron transistor (SET). The trap is connected to
the box through a highly resistive tunnel junction. To measure the
box charge state, the trap is biased by a readout pulse applied to
the readout gate, so that if the box is in the excited state, an
extra Cooper-pair charge tunnels into the trap in a sequential
two-quasiparticle process and then detected by the SET. Note that
the quantum states are not much decohered by the measurement
circuit until the readout pulse is applied, as the SET is
effectively decoupled from the qubit by serial capacitive
dividers. Mutual and self capacitances of the islands are
designated by $C_{ij}$ and $C_{i}$, where $i$ and $j$ are
characters $b$, $t$, and $s$, denoting the box, trap and SET
islands, respectively. We have measured values of these
capacitances ($C_{b} \approx$ 600 aF, $C_{t}\approx C_{s} \approx$
1000 aF, $C_{bt}\approx 30$ aF, $C_{st} \approx$ 100 aF) and also
gate capacitances to their nearest islands. Other stray
capacitances between the box and remote elements (between the box
and each existing lead including the SET island and the ones not
drawn in Fig.~\ref{fig:T1cFig1}(a)) are calculated to be less than
0.3 aF. Electric field of those elements is strongly screened by
the 100 nm thick Au ground plane beneath the 400 nm thick
Si$_{3}$N$_{4}$ insulating layer. The box charging energy is
$E_{C} \approx e^{2}/2C_{b} =$ 130 $\mu$eV ($E_{C}/h$ = 32 GHz).
The reservoir is a big island galvanically isolated from
electrical leads and has capacitance of about 0.1 nF to the ground
plane. The SET with 200-k$\Omega$ junction resistances is usually
biased at the Josephson quasiparticle cycle peak \cite{Fulton}
with the maximum current of about 100 pA.

In the charge basis $|0\rangle$ and $|1\rangle$ (without and with
an extra Cooper-pair in the box), the Hamiltonian of the box can
be written as
\begin{equation}
H = - \frac{\Delta E}{2} [\sigma_{z} \cos\theta + \sigma_{x}
\sin\theta],\label{eq:Eq1}
\end{equation}
where $\sigma_{z}$, $\sigma_{x}$ are the Pauli matrices, $\Delta E
= (U^{2} + E_{J}^{2})^{1/2}$, $U = 2eq/C_{b}$ is the electrostatic
energy difference between the two states, $q$ is a gate induced
charge ($q \equiv$ 0 at the degeneracy point) and $\theta = -
\arctan(E_{J}/U)$. Eigenstates of the two-level system are
$|+\rangle = \cos (\theta/2) |0\rangle + \sin (\theta/2)
|1\rangle$ and $|-\rangle = - \sin (\theta/2) |0\rangle + \cos
(\theta/2) |1\rangle$ with corresponding eigenenergies $- \Delta
E/2$ and $\Delta E/2$. Fig.~\ref{fig:T1cFig1}(b) schematically
shows the energy diagram of the qubit as a function of $q$. Solid
and dashed lines represent eigenenergies and electrostatic
energies, respectively. We usually adjust the qubit to a position
of $q_{0}$ ($< 0$), where $|U| \gg E_{J}$ and eigenstates are
nearly pure charge states $|+\rangle \approx |0\rangle$ and
$|-\rangle \approx |1\rangle$. For the coherent control of the
qubit, we start from the ground state $|0\rangle$ and apply a
rectangular pulse bringing the system nonadiabatically to the
degeneracy point ($q = 0$, $\theta = \pi/2$) for the time $t_{p}$,
where the state freely evolves as $\cos (E_{J}t/2\hbar) |0\rangle
+ i \sin (E_{J}t/2\hbar) |1\rangle$. We create state $|1\rangle$
by applying the pulse of length $t_{p} = 2\pi\hbar/E_{J}$
($\pi$-pulse) \cite{Squbit}.

To measure energy relaxation dynamics of the excited state
$|-\rangle$  we use a combination of the $\pi$-pulse and an
adiabatic pulse (a pulse with slow rise and fall times satisfying
the condition of $\hbar|d\Delta E/dt| \ll E_{J}^{2}$). The
manipulation procedure, schematically shown in
Fig.~\ref{fig:T1cFig1}(b), includes three sequential steps: First,
the $\pi$-pulse is applied to the box to prepare the excited state
$|1\rangle$. Second, an adiabatic pulse is applied to the box, so
that its rise front shifts the system along the excited state
$|-\rangle$ to a point $q = q_{0} + q_{a}$ and holds the system at
fixed $q$ for a time $t_{a}$, where relaxation from the excited
state $|-\rangle$ to the ground state $|+\rangle$ may occur with a
finite probability dependent on $t_{a}$. Third, the fall of the
adiabatic pulse converts the excited state $|-\rangle$ to the
charge state $|1\rangle$ and the ground state $|+\rangle$ to state
$|0\rangle$, respectively. One can study the relaxation dynamics
at a desired value of $q$ by measuring probability $P$ of
detecting the excited state $|1\rangle$ in the final state as a
function of $t_{a}$.

\begin{figure}
\includegraphics{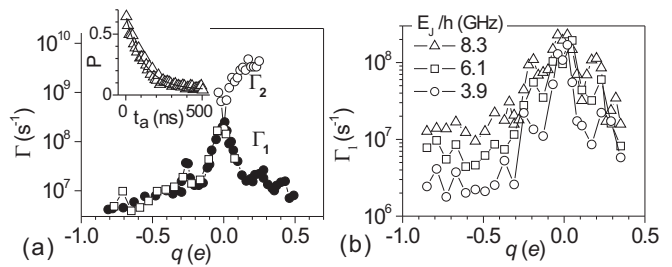}
\caption{\label{fig:T1cFig2} (a) Energy relaxation rate
$\Gamma_{1}$ measured by the SET normally set in the JQP peak
(closed circles) and in blockade regime (open squares) and phase
decoherence rate $\Gamma_{2}$ (open circles) versus gate induced
charge $q$ of sample I ($E_{J}/h$ = 5.1 GHz). The inset
exemplifies a decay of probability $P$ to detect charge in the
trap as a function of the adiabatic pulse length $t_{a}$, measured
at $q = -0.36$ $e$. $\Gamma_{1}$ is derived from exponential fit
of the decay. (b) $\Gamma_{1}$ measured in sample II for three
different Josephson energies.}
\end{figure}

We study two samples (I and II) of an identical geometry at a
temperature of 50 mK. Probability $P$ is determined by repeating
nominally identical quantum state manipulations and readouts
\cite{RComm}. The inset of Fig.~\ref{fig:T1cFig2}(a) exemplifies a
typical decay of $P$ as a function of $t_{a}$ at $q = - 0.36$ $e$
of sample I. We derive the energy relaxation rate $\Gamma_{1}$ by
fitting the data with $A \exp(-\Gamma_{1} t_{a}) + B$ using three
fitting parameters $A$, $B$ and $\Gamma_{1}$. The amplitude $A$
depends on the efficiency of each step of the state manipulations
and is independent of $t_{a}$. $A$ is a constant at fixed $q$,
when all parameters except $t_{a}$ are kept unchanged. A small
finite value of $B$ is a consequence of ``dark" switches in our
circuit. Note that in the present experiments $B$ is always small
independently of $q$, indicating that relaxation is much stronger
than excitation.

\begin{figure}
\includegraphics{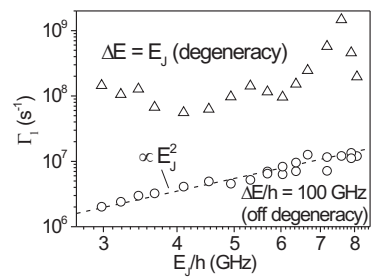}
\caption{\label{fig:T1cFig3} Relaxation rate  $\Gamma_{1}$ as a
function of $E_{J}$ measured at the degeneracy point (open
triangles) and off the degeneracy point (open circles) at $\Delta
E/h$ $\approx$ 100 GHz for sample I.}
\end{figure}

Figs.~\ref{fig:T1cFig2} (a), (b) show $\Gamma_{1}$ as a function
of $q$. Fig.~\ref{fig:T1cFig2}(a) demonstrates $\Gamma_{1}$
(closed circles) measured in sample I with $E_{J}/h$ = 5.1 GHz.
Fig.~\ref{fig:T1cFig2}(b) shows $\Gamma_{1}$ in sample II measured
for three different Josephson energies: $E_{J}/h$ = 3.9 GHz
(dots), 6.1 GHz (squares) and 8.3 GHz (triangles). $E_{J}$
dependences of the relaxation rate $\Gamma_{1}$ for sample II at
fixed adiabatic pulse amplitudes are shown in
Fig.~\ref{fig:T1cFig3}. $\Gamma_{1}$ at $q = -0.8$ $e$ ($\Delta
E/h$ = 100 GHz) is presented by open circles, while $\Gamma_{1}$
at $q = 0$ (degeneracy point) is presented by open triangles.

For the charge qubit, noises coupled to the qubit charge degree of
freedom are presumably the main origin of decoherence. If the
energy fluctuations of the qubit due to the charge noise are
characterized by the spectral density of electrostatic energy
fluctuations $S_{U}(\omega)$ \cite{Sw}, the relaxation rate is
given by the Fermi's golden rule (see e.g. Ref.
\cite{Schoelkopf2})
\begin{equation}
\Gamma_{1} = \frac{\pi S_{U}(\omega)}{2\hbar^{2}} \sin^{2}\theta.
\label{eq:Eq2}
\end{equation}
The overall behavior of  $\Gamma_{1}$ in
Figs.~\ref{fig:T1cFig2}(a), (b) --- $\Gamma_{1}$ decreases with
increasing $|q|$ --- is described by $\sin^{2}\theta =
E_{J}^{2}/(E_{J}^{2}+U^{2})$, characterizing coupling of the qubit
to the reservoir via the charge degree of freedom. When $\Delta E
\gg E_{J}$, Eq.~(\ref{eq:Eq2}) can be rewritten in the form
$\Gamma_{1} \approx \pi S_{U} (\omega \approx U/\hbar)
(E_{J}/U)^{2}/(2\hbar^{2})$. The experimentally measured
$\Gamma_{1}$ at $\Delta E/h$ = 100 GHz (open circles in
Fig.~\ref{fig:T1cFig3}) has clear $E_{J}^{2}$-dependence (dashed
curve). This confirms the dominant contribution of charge
fluctuations. Moreover, it indicates that this relaxation process
is Cooper-pair tunneling rather than sequential tunneling of two
quasiparticles. At the degeneracy point, $\Gamma_{1} = \pi S_{U}
(\omega = E_{J}/\hbar)/2\hbar^{2}$ (open triangles in
Fig.~\ref{fig:T1cFig3}) directly reproduces frequency dependence
of $S_{U}(\omega)$.

It is known that one of the main low-frequency noises in
nano-scale charge devices is the $1/f$ noise produced by a bath of
charge fluctuators. The spectral density of the charge
fluctuations defined for negative and positive frequencies is
given by
\begin{equation}
S_{q}(\omega) = \frac{\alpha}{2|\omega|}. \label{eq:Eq3}
\end{equation}
The parameter $\alpha$ has been found in a number of experiments
to be typically of order of (10$^{-3} e)^{2}$
\cite{Noise1,Mooij,PTB,Dephasing}. It was also shown in charge
echo experiments that the $1/f$ noise reasonably explains
dephasing of the charge qubit \cite{Dephasing}.

Under an assumption of the Gaussian noise produced by many
fluctuators weakly coupled to the qubit, the coherent oscillations
dephase as $\exp[-\varphi(t)]$, as the random phase
\begin{equation}
\varphi(t)\approx\frac{\cos^{2}\theta}{\hbar^{2}}\int_{\omega_{0}}^{\infty}S_{U}(\omega)\biggl[
\frac{2\sin(\omega t/2)}{\omega}\biggr]^{2}d\omega \label{eq:Eq4}
\end{equation}
is accumulated from the low-frequency component of $S_{U}(\omega)$
($|\omega| \leq 2\pi/t$), where $S_{U}(\omega) =
(4E_{C}/e)^{2}S_{q}(\omega)$. The phase decoherence time $T_{2} =
\Gamma_{2}^{-1}$ is defined as $\varphi(T_{2}) = 1$. From the
experimentally observed decay of coherent oscillations,
$\Gamma_{2}$ is obtained for various $q$ in sample I
(Fig.~\ref{fig:T1cFig2} (a)). Assuming the $1/f$ noise spectrum of
Eq.~(\ref{eq:Eq3}), we obtain the parameter $\alpha = [\eta \hbar
e \Gamma_{2}/(E_{c}\cos\theta)]^{2} \approx
(1.3\times10^{-3}e)^{2}$, where $\eta$ is a numeric coefficient
weakly dependent on the lower cutoff frequency $\omega_{0}$
\cite{Eta}.

\begin{figure}
\includegraphics{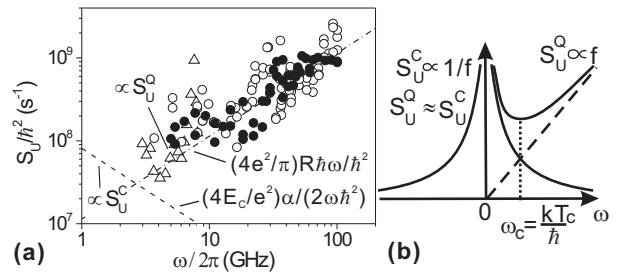}
\caption{\label{fig:T1cFig4}(a) Noise $S_{U}/\hbar^{2}$ derived
from $\Gamma_{1} - q$ dependences of the sample I (closed circles)
and sample II (open circles). Open triangles show $\Gamma_{1}$
dependences of sample II at the degeneracy point. A dashed-dotted
line represents a linear rise $(4e^{2}/\pi)R\hbar\omega/\hbar^{2}$
($f$ noise) with $R = 6$~$\Omega$. A dashed line is the $1/f$
noise from Eq. (\ref{eq:Eq3}) with $\alpha \approx
(1.3\times10^{-3} e)^{2}$ derived from $\Gamma_{2}$ measurements.
(b) A schematic representation of the classical
$S^{C}_{U}(\omega)$ and quantum $S^{Q}_{U}(\omega)$ noise
behavior. At frequencies $\omega < kT/\hbar$,
$S^{Q}_{U}(\omega)\approx S^{C}_{U}(\omega)$ and behaves like a
$1/f$ noise. At frequencies $|\omega| > kT/\hbar$,
$S^{Q}_{U}(\omega)
> S^{C}_{U}(\omega)$ and the quantum noise is proportional to
$\omega$.}
\end{figure}

Fig.~\ref{fig:T1cFig4}(a) summarizes reduced noise spectra
$S_{U}/\hbar^{2} = 2\Gamma_{1}/(\pi\sin^{2}\theta)$ derived from
$\Gamma_{1}$ mesurements. $S_{U}/\hbar^{2}$ derived from
$\Gamma_{1} - q$ dependences in sample I is plotted by closed
circles, while $S_{U}/\hbar^{2}$ for sample II with different
$E_{J}$ is plotted by open circles. In addition, $S_{U}/\hbar^{2}$
= $2\Gamma_{1}/\pi$ measured at the degeneracy point of sample II
is plotted by open triangles. We also show $S_{U}/\hbar^{2}$ for
the $1/f$ noise (Eq.~(\ref{eq:Eq3})) with $\alpha =
(1.3\times10^{-3} e)^{2}$ by the dashed line. The data exhibits
rise with increasing $\omega$. The dashed-dotted line exemplifies
linear dependence (as in the case of ohmic environment), which we
present in the form of $(4e^{2}/\pi)R\hbar\omega/\hbar^{2}$ with
$R = 6$~$\Omega$. The actual rise of the experimental data is not
monotonic but has some resonance-like peaks, for instance, at 7
and 30 GHz. This probably reflects coupling to some resonances,
which can be other microscopic coherent two-level systems or
geometrical resonances in the device.
The crossover frequency of the $1/f$ and $f$ curves is $\omega_{c}
= 2\pi\times2.6$ GHz, which formally corresponds to the
temperature $T_{c} = \hbar \omega_{c}/k$ = 120 mK.

We argue that in our device, the relaxation cannot be explained by
coupling to electromagnetic environment through electrical leads
of the gates and the readout circuit. The fact that the qubit
relaxation is much stronger than excitation indicates that the
relaxation is induced neither by radiation from hot environment
nor by power dissipated in the measurement SET. For spontaneous
emission to the environment, the lead impedance needed to explain
$\Gamma_1$ is estimated. The control-gate-to-box capacitance
$C_{cg}=1$ aF gives coupling strength $\kappa_g = C_{cg}/C_b
\approx 2 \times 10^{-3} $. Other leads have direct capacitances
to the box smaller than 0.3~aF. Even for the indirect coupling to
the box through the trap, the strongest coupling factor is
$\kappa_g \approx C_{rg} C_{bt}/ C_b C_t \approx 5 \times 10
^{-3}$ to the readout gate (readout-gate-to-trap capacitance
$C_{rg} \approx 10$~aF). The lead impedance required to explain
$\Gamma_1$ is $R/\kappa_g^2 \sim 10^5$~$\Omega$ and much higher
than the typical value of $\sim$ 100~$\Omega$. Also, $\Gamma_1$
can not be explained by absorption on the SET. The box-to-SET
coupling is $\kappa_{s} \approx C_{st}C_{bt}/C_{b}C_{t} \approx
5\times10^{-3}$ and SET impedance is substituted by $Z_{s}$ = $
R_{s}/(1 + i\omega C_{s} R_{s})$, where $R_s \geq 100$~k$\Omega$
depends on the SET gate bias conditions with minimal value of two
parallel tunnel junction resistances. The dissipative impedance
for our qubit is $\kappa_{s}^{2}$Re[$Z_{s}$] $\leq$ 1.6~$\Omega$
at $\omega$ = 5 GHz and rapidly decreases with frequency and
resistance as $\omega^{-2}R_s^{-1}$.
To check experimentally the effect of the SET
noise, we compare in Fig.~2(a) $\Gamma_1$ obtained with SET biased
at the JQP peak (closed circles) and in the blockade regime (open
squares; SET current switches on only when charge is trapped).
Independence of $\Gamma_1$ of bias conditions proves that the SET
noise is not dominant in the qubit relaxation.

The observed $E_J^2$-dependence shows that the dominant noise is
coupled to the charge degree of freedom. However, with the
discussions in the previous paragraph, we exclude a possible
contribution of the coupling to the electromagnetic environment
through the leads. On the other hand, dephasing of our qubit is
well explained by the 1/$f$ noise produced by charge fluctuators.
In standard models of the 1/$f$ noise, the fluctuators are
characterized by activation energy $E^*$ and switching rate
$\gamma$ \cite{1fnoise}. In thermal equilibrium, the charge
fluctuators with exciation energies $E^* \leq kT$ are activated
producing telegraph-like noise and contributing to the 1/$f$ noise
at frequency $\gamma$. However, we suppose that inactive
fluctuators with $E^* \geq kT$ may absorb energy $\Delta E$ of our
qubit, which is out of equilibrium with surrounding environment
($\Delta E > kT$).
The thermally excited fluctuators produce the classical noise
(without energy exchange with the qubit) $S_{U}^{C}$
($S_{U}^{C}(\omega) = S_{U}^{C}(-\omega)$) and cause the qubit
dephasing, while the spontaneous emission is characterized by the
quantum noise $S_{U}^{Q}$ ($S_{U}^{Q}(|\omega|)>
S_{U}^{Q}(-|\omega|)$). Namely, the ``hot" qubit ($\Delta E > kT$)
may only release the excess energy $\Delta E$ to the ``cold" bath.
Fig.~\ref{fig:T1cFig4}(b) schematically represents the noise
behavior extended to the negative frequency range. $S_{U}^{Q}$
almost coincides with $S_{U}^{C}$ at low frequencies $\hbar \omega
\ll kT$ (``hot" bath), where the qubit has an equal chance to emit
and to absorb energy. $S_{U}^{C}$ is expected to have 1/$f$
dependence, while $S_{U}^{Q}$ at $\hbar \omega \gg kT$ (``cold"
bath) is roughly proportional to $f$ according to our experimental
data.

The crossover frequency $\omega_{c}$ satisfies the condition
$\alpha/2\omega_{c} = \beta \hbar \omega_{c}$, where $\beta \hbar
\omega$ is the $f$-noise characterized by temperature independent
parameter $\beta$. The corresponding 1/$f$ noise spectral density
at $|\omega|\ll |\omega_{c}|$ is
\begin{equation}
S_{U}^{C}(\omega) = \frac{\beta(kT_{c})^2}{\hbar|\omega|}.
\label{eq:Eq5}
\end{equation}
This means that if $T_{c}$ scales with surrounding temperature
$T$, then the 1/$f$ noise should have $T^{2}$ dependence. For
instance, if the density of states of the fluctuators is constant
or a weak function of two independent energies (for example,
excitation energies of an electron in a double-well potential)
then the $1/f$ noise is proportional to $T^{2}$ \cite{Kenyon}. The
temperature dependence of the 1/$f$-noise can be verified
experimentally by measuring for example SET low frequency noise.


In conclusion, we have measured energy relaxation and dephasing of
the Josephson charge qubit. The relaxation is a spontaneous
emission process induced by the quantum charge noise nearly
proportional to the qubit energy ($f$ quantum noise). Dephasing of
the qubit is explained by the 1/$f$ noise. The $f$ quantum noise
crosses over the extrapolated $1/f$ classical noise (causing pure
dephasing) at a frequency $\omega_{c} = kT_{c}/\hbar$, with $T_{c}
\approx$ 120 mK.

We thank B. Altshuler, D. Averin, P. Delsing, D. Estive, G. Falci,
S. Lloyd, E. Paladino, A. Shnirman and D. Vion for valuable
discussions.


\begin{thebibliography}{23}
\expandafter\ifx\csname
natexlab\endcsname\relax\def\natexlab#1{#1}\fi
\expandafter\ifx\csname bibnamefont\endcsname\relax
  \def\bibnamefont#1{#1}\fi
\expandafter\ifx\csname bibfnamefont\endcsname\relax
  \def\bibfnamefont#1{#1}\fi
\expandafter\ifx\csname citenamefont\endcsname\relax
  \def\citenamefont#1{#1}\fi
\expandafter\ifx\csname url\endcsname\relax
  \def\url#1{\texttt{#1}}\fi
\expandafter\ifx\csname
urlprefix\endcsname\relax\def\urlprefix{URL }\fi
\providecommand{\bibinfo}[2]{#2}
\providecommand{\eprint}[2][]{\url{#2}}

\bibitem[{\citenamefont{Nakamura et~al.}(1999)\citenamefont{Nakamura, Pashkin,
  and Tsai}}]{Squbit}
\bibinfo{author}{\bibfnamefont{Y.}~\bibnamefont{Nakamura}},
  \bibinfo{author}{\bibfnamefont{Y.~A.} \bibnamefont{Pashkin}},
  \bibnamefont{and} \bibinfo{author}{\bibfnamefont{J.~S.} \bibnamefont{Tsai}},
  \bibinfo{journal}{Nature} \textbf{\bibinfo{volume}{398}},
  \bibinfo{pages}{786} (\bibinfo{year}{1999}).

\bibitem[{\citenamefont{Vion et~al.}(2002)\citenamefont{Vion, Aassime, Cottet,
  Joyez, Pothier, Urbina, Esteve, and Devoret}}]{Saclay}
\bibinfo{author}{\bibfnamefont{D.}~\bibnamefont{Vion}},
  \bibinfo{author}{\bibfnamefont{A.}~\bibnamefont{Aassime}},
  \bibinfo{author}{\bibfnamefont{A.}~\bibnamefont{Cottet}},
  \bibinfo{author}{\bibfnamefont{P.}~\bibnamefont{Joyez}},
  \bibinfo{author}{\bibfnamefont{H.}~\bibnamefont{Pothier}},
  \bibinfo{author}{\bibfnamefont{C.}~\bibnamefont{Urbina}},
  \bibinfo{author}{\bibfnamefont{D.}~\bibnamefont{Esteve}}, \bibnamefont{and}
  \bibinfo{author}{\bibfnamefont{M.~H.} \bibnamefont{Devoret}},
  \bibinfo{journal}{Science} \textbf{\bibinfo{volume}{296}},
  \bibinfo{pages}{886} (\bibinfo{year}{2002}).

\bibitem[{\citenamefont{Yu et~al.}(2002)\citenamefont{Yu, Han, Chu, Chu, and
  Wang}}]{Yu}
\bibinfo{author}{\bibfnamefont{Y.}~\bibnamefont{Yu}},
  \bibinfo{author}{\bibfnamefont{S.}~\bibnamefont{Han}},
  \bibinfo{author}{\bibfnamefont{X.}~\bibnamefont{Chu}},
  \bibinfo{author}{\bibfnamefont{S.~I.} \bibnamefont{Chu}}, \bibnamefont{and}
  \bibinfo{author}{\bibfnamefont{Z.}~\bibnamefont{Wang}},
  \bibinfo{journal}{Science} \textbf{\bibinfo{volume}{296}},
  \bibinfo{pages}{889} (\bibinfo{year}{2002}).

\bibitem[{\citenamefont{Martinis et~al.}(2002)\citenamefont{Martinis, Nam,
  Aumentado, and Urbina}}]{Martinis1}
\bibinfo{author}{\bibfnamefont{J.~M.} \bibnamefont{Martinis}},
  \bibinfo{author}{\bibfnamefont{S.}~\bibnamefont{Nam}},
  \bibinfo{author}{\bibfnamefont{J.}~\bibnamefont{Aumentado}},
  \bibnamefont{and} \bibinfo{author}{\bibfnamefont{C.}~\bibnamefont{Urbina}},
  \bibinfo{journal}{Phys.\ Rev.\ Lett.} \textbf{\bibinfo{volume}{89}},
  \bibinfo{pages}{117901} (\bibinfo{year}{2002}).

\bibitem[{\citenamefont{Chiorescu et~al.}(2003)\citenamefont{Chiorescu,
  Nakamura, Harmans, and Mooij}}]{Delft}
\bibinfo{author}{\bibfnamefont{I.}~\bibnamefont{Chiorescu}},
  \bibinfo{author}{\bibfnamefont{Y.}~\bibnamefont{Nakamura}},
  \bibinfo{author}{\bibfnamefont{C.~J. P.~M.} \bibnamefont{Harmans}},
  \bibnamefont{and} \bibinfo{author}{\bibfnamefont{J.~E.} \bibnamefont{Mooij}},
  \bibinfo{journal}{Science} \textbf{\bibinfo{volume}{299}},
  \bibinfo{pages}{1869} (\bibinfo{year}{2003}).

\bibitem[{\citenamefont{Duty et~al.}(2004)\citenamefont{Duty, Gunnarsson,
  Bladh, and Delsing}}]{Chalmers}
\bibinfo{author}{\bibfnamefont{T.}~\bibnamefont{Duty}},
  \bibinfo{author}{\bibfnamefont{D.}~\bibnamefont{Gunnarsson}},
  \bibinfo{author}{\bibfnamefont{K.}~\bibnamefont{Bladh}}, \bibnamefont{and}
  \bibinfo{author}{\bibfnamefont{P.}~\bibnamefont{Delsing}},
  \bibinfo{journal}{Phys.\ Rev. B} \textbf{\bibinfo{volume}{69}},
  \bibinfo{pages}{140503} (\bibinfo{year}{2004}).

\bibitem[{\citenamefont{Pashkin et~al.}(2003)\citenamefont{Pashkin, Yamamoto,
  Astafiev, Nakamura, Averin, and Tsai}}]{Dqubit}
\bibinfo{author}{\bibfnamefont{Y.~A.} \bibnamefont{Pashkin}},
  \bibinfo{author}{\bibfnamefont{T.}~\bibnamefont{Yamamoto}},
  \bibinfo{author}{\bibfnamefont{O.}~\bibnamefont{Astafiev}},
  \bibinfo{author}{\bibfnamefont{Y.}~\bibnamefont{Nakamura}},
  \bibinfo{author}{\bibfnamefont{D.~V.} \bibnamefont{Averin}},
  \bibnamefont{and} \bibinfo{author}{\bibfnamefont{J.~S.} \bibnamefont{Tsai}},
  \bibinfo{journal}{Nature} \textbf{\bibinfo{volume}{421}},
  \bibinfo{pages}{823} (\bibinfo{year}{2003}).

\bibitem[{\citenamefont{Yamamoto et~al.}(2003)\citenamefont{Yamamoto, Pashkin,
  Astafiev, Nakamura, and Tsai}}]{CNOT}
\bibinfo{author}{\bibfnamefont{T.}~\bibnamefont{Yamamoto}},
  \bibinfo{author}{\bibfnamefont{Y.~A.} \bibnamefont{Pashkin}},
  \bibinfo{author}{\bibfnamefont{O.}~\bibnamefont{Astafiev}},
  \bibinfo{author}{\bibfnamefont{Y.}~\bibnamefont{Nakamura}}, \bibnamefont{and}
  \bibinfo{author}{\bibfnamefont{J.~S.} \bibnamefont{Tsai}},
  \bibinfo{journal}{Nature} \textbf{\bibinfo{volume}{425}},
  \bibinfo{pages}{941} (\bibinfo{year}{2003}).

\bibitem[{\citenamefont{Makhlin et~al.}(2001)\citenamefont{Makhlin,
  Sch{\"{o}}n, and Shnirman}}]{Makhlin}
\bibinfo{author}{\bibfnamefont{Y.}~\bibnamefont{Makhlin}},
  \bibinfo{author}{\bibfnamefont{G.}~\bibnamefont{Sch{\"{o}}n}},
  \bibnamefont{and} \bibinfo{author}{\bibfnamefont{A.}~\bibnamefont{Shnirman}},
  \bibinfo{journal}{Rev.\ Mod.\ Phys.} \textbf{\bibinfo{volume}{73}},
  \bibinfo{pages}{357} (\bibinfo{year}{2001}).

\bibitem[{\citenamefont{Averin}(2000)}]{Averin}
\bibinfo{author}{\bibfnamefont{D.~V.} \bibnamefont{Averin}},
  \bibinfo{journal}{Fortshr.\ Phys.} \textbf{\bibinfo{volume}{48}},
  \bibinfo{pages}{1055} (\bibinfo{year}{2000}).

\bibitem[{\citenamefont{Nakamura et~al.}(2002)\citenamefont{Nakamura, Pashkin,
  Yamamoto, and Tsai}}]{Dephasing}
\bibinfo{author}{\bibfnamefont{Y.}~\bibnamefont{Nakamura}},
  \bibinfo{author}{\bibfnamefont{Y.~A.} \bibnamefont{Pashkin}},
  \bibinfo{author}{\bibfnamefont{T.}~\bibnamefont{Yamamoto}}, \bibnamefont{and}
  \bibinfo{author}{\bibfnamefont{J.~S.} \bibnamefont{Tsai}},
  \bibinfo{journal}{Phys.\ Rev.\ Lett.} \textbf{\bibinfo{volume}{88}},
  \bibinfo{pages}{047901} (\bibinfo{year}{2002}).

\bibitem[{\citenamefont{Lehnert et~al.}(2003)\citenamefont{Lehnert, Bladh,
  Spietz, Gunnarson, Schuster, Delsing, and Schoelkopf}}]{Schoelkopf}
\bibinfo{author}{\bibfnamefont{K.~W.} \bibnamefont{Lehnert}},
  \bibinfo{author}{\bibfnamefont{K.}~\bibnamefont{Bladh}},
  \bibinfo{author}{\bibfnamefont{L.~F.} \bibnamefont{Spietz}},
  \bibinfo{author}{\bibfnamefont{D.}~\bibnamefont{Gunnarson}},
  \bibinfo{author}{\bibfnamefont{D.~I.} \bibnamefont{Schuster}},
  \bibinfo{author}{\bibfnamefont{P.}~\bibnamefont{Delsing}}, \bibnamefont{and}
  \bibinfo{author}{\bibfnamefont{R.~J.} \bibnamefont{Schoelkopf}},
  \bibinfo{journal}{Phys.\ Rev.\ Lett.} \textbf{\bibinfo{volume}{90}},
  \bibinfo{pages}{027002} (\bibinfo{year}{2003}).

\bibitem[{\citenamefont{Vion et~al.}(2003)\citenamefont{Vion, Aassime, Cottet,
  Joyez, Pothier, Urbina, Esteve, and Devoret}}]{Decsac}
\bibinfo{author}{\bibfnamefont{D.}~\bibnamefont{Vion}},
  \bibinfo{author}{\bibfnamefont{A.}~\bibnamefont{Aassime}},
  \bibinfo{author}{\bibfnamefont{A.}~\bibnamefont{Cottet}},
  \bibinfo{author}{\bibfnamefont{P.}~\bibnamefont{Joyez}},
  \bibinfo{author}{\bibfnamefont{H.}~\bibnamefont{Pothier}},
  \bibinfo{author}{\bibfnamefont{C.}~\bibnamefont{Urbina}},
  \bibinfo{author}{\bibfnamefont{D.}~\bibnamefont{Esteve}}, \bibnamefont{and}
  \bibinfo{author}{\bibfnamefont{M.}~\bibnamefont{Devoret}},
  \bibinfo{journal}{Fortshr.\ Phys.} \textbf{\bibinfo{volume}{51}},
  \bibinfo{pages}{462} (\bibinfo{year}{2003}).

\bibitem[{\citenamefont{Schoelkopf et~al.}(2002)\citenamefont{Schoelkopf,
  Clerk, Girvin, Lehnert, and Devoret}}]{Schoelkopf2}
\bibinfo{author}{\bibfnamefont{R.~J.} \bibnamefont{Schoelkopf}},
  \bibinfo{author}{\bibfnamefont{A.~A.} \bibnamefont{Clerk}},
  \bibinfo{author}{\bibfnamefont{S.~M.} \bibnamefont{Girvin}},
  \bibinfo{author}{\bibfnamefont{K.~W.} \bibnamefont{Lehnert}},
  \bibnamefont{and} \bibinfo{author}{\bibfnamefont{M.}~\bibnamefont{Devoret}},
  in \emph{\bibinfo{booktitle}{Quantum noise in mesoscopic physics.}}, edited
  by \bibinfo{editor}{\bibfnamefont{Y.~V.} \bibnamefont{Nazarov}}
  (\bibinfo{publisher}{Kluwer, Dordrecht}, \bibinfo{year}{2002}), pp.
  \bibinfo{pages}{175--203}.

\bibitem[{\citenamefont{Astafiev et~al.}(2004)\citenamefont{Astafiev, Pashkin,
  Yamamoto, Nakamura, and Tsai}}]{RComm}
\bibinfo{author}{\bibfnamefont{O.}~\bibnamefont{Astafiev}},
  \bibinfo{author}{\bibfnamefont{Y.~A.} \bibnamefont{Pashkin}},
  \bibinfo{author}{\bibfnamefont{T.}~\bibnamefont{Yamamoto}},
  \bibinfo{author}{\bibfnamefont{Y.}~\bibnamefont{Nakamura}}, \bibnamefont{and}
  \bibinfo{author}{\bibfnamefont{J.~S.} \bibnamefont{Tsai}},
  \bibinfo{journal}{Phys.\ Rev. B} \textbf{\bibinfo{volume}{69}},
  \bibinfo{pages}{180507} (\bibinfo{year}{2004}).

\bibitem[{\citenamefont{Fulton et~al.}(1989)\citenamefont{Fulton, Gammel,
  Bishop, Dunkleberger, and Dolan}}]{Fulton}
\bibinfo{author}{\bibfnamefont{T.~A.} \bibnamefont{Fulton}},
  \bibinfo{author}{\bibfnamefont{P.~L.} \bibnamefont{Gammel}},
  \bibinfo{author}{\bibfnamefont{D.~J.} \bibnamefont{Bishop}},
  \bibinfo{author}{\bibfnamefont{L.~N.} \bibnamefont{Dunkleberger}},
  \bibnamefont{and} \bibinfo{author}{\bibfnamefont{G.~J.} \bibnamefont{Dolan}},
  \bibinfo{journal}{Phys.\ Rev.\ Lett.} \textbf{\bibinfo{volume}{63}},
  \bibinfo{pages}{1307} (\bibinfo{year}{1989}).

\bibitem[{Sw()}]{Sw}
\bibinfo{note}{Here, we define the spectral density for operator
  $\widehat{g}(t)$ as $S_{g}(\omega)=
  \frac{1}{2\pi}\int_{-\infty}^{\infty}{\langle\widehat{g}(\tau)\widehat{g}(0)%
}\rangle e^{-i\omega\tau} d\tau$, where angular brackets represent
quantum
  statistical averaging}.

\bibitem[{\citenamefont{Zimmerli et~al.}(1992)\citenamefont{Zimmerli, Eiles,
  Kautz, and Martinis}}]{Noise1}
\bibinfo{author}{\bibfnamefont{G.}~\bibnamefont{Zimmerli}},
  \bibinfo{author}{\bibfnamefont{T.~M.} \bibnamefont{Eiles}},
  \bibinfo{author}{\bibfnamefont{R.~L.} \bibnamefont{Kautz}}, \bibnamefont{and}
  \bibinfo{author}{\bibfnamefont{J.~M.} \bibnamefont{Martinis}},
  \bibinfo{journal}{Appl.\ Phys.\ Lett.} \textbf{\bibinfo{volume}{61}},
  \bibinfo{pages}{237} (\bibinfo{year}{1992}).

\bibitem[{\citenamefont{Verbrugh et~al.}(1995)\citenamefont{Verbrugh,
  Benhamadi, Visscher, and Mooij}}]{Mooij}
\bibinfo{author}{\bibfnamefont{S.~M.} \bibnamefont{Verbrugh}},
  \bibinfo{author}{\bibfnamefont{M.~L.} \bibnamefont{Benhamadi}},
  \bibinfo{author}{\bibfnamefont{E.~H.} \bibnamefont{Visscher}},
  \bibnamefont{and} \bibinfo{author}{\bibfnamefont{J.~E.} \bibnamefont{Mooij}},
  \bibinfo{journal}{J. Appl.\ Phys.} \textbf{\bibinfo{volume}{78}},
  \bibinfo{pages}{2830} (\bibinfo{year}{1995}).

\bibitem[{\citenamefont{Wolf et~al.}(1997)\citenamefont{Wolf, Ahlers, Niemeyer,
  Scherer, Weimann, Zorin, Krupenin, Lotkhov, and Presnov}}]{PTB}
\bibinfo{author}{\bibfnamefont{H.}~\bibnamefont{Wolf}},
  \bibinfo{author}{\bibfnamefont{F.~J.} \bibnamefont{Ahlers}},
  \bibinfo{author}{\bibfnamefont{J.}~\bibnamefont{Niemeyer}},
  \bibinfo{author}{\bibfnamefont{H.}~\bibnamefont{Scherer}},
  \bibinfo{author}{\bibfnamefont{T.}~\bibnamefont{Weimann}},
  \bibinfo{author}{\bibfnamefont{A.~B.} \bibnamefont{Zorin}},
  \bibinfo{author}{\bibfnamefont{V.~A.} \bibnamefont{Krupenin}},
  \bibinfo{author}{\bibfnamefont{S.~V.} \bibnamefont{Lotkhov}},
  \bibnamefont{and} \bibinfo{author}{\bibfnamefont{D.~E.}
  \bibnamefont{Presnov}}, \bibinfo{journal}{IEEE Trans. Instr. Meas.}
  \textbf{\bibinfo{volume}{46}}, \bibinfo{pages}{303} (\bibinfo{year}{1997}).

\bibitem[{Eta()}]{Eta}
\bibinfo{note}{Taking $\omega_{0} = \pi/\tau_{m}$ with a typical measurement
  time of one data point $\tau_{m} = 0.6$ s, we find $\eta \approx 0.055$.}

\bibitem[{\citenamefont{Kogan}(1996)}]{1fnoise}
\bibinfo{author}{\bibfnamefont{S.}~\bibnamefont{Kogan}},
  \emph{\bibinfo{title}{Electronic noise and fluctuators in solids}}
  (\bibinfo{publisher}{Cambridge University Press},
  \bibinfo{address}{Cambridge}, \bibinfo{year}{1996}).

\bibitem[{\citenamefont{Kenyon et~al.}(2000)\citenamefont{Kenyon, Lobb, and
  Wellstood}}]{Kenyon}
\bibinfo{author}{\bibfnamefont{M.}~\bibnamefont{Kenyon}},
  \bibinfo{author}{\bibfnamefont{C.~J.} \bibnamefont{Lobb}}, \bibnamefont{and}
  \bibinfo{author}{\bibfnamefont{F.~C.} \bibnamefont{Wellstood}},
  \bibinfo{journal}{J.\ Appl.\ Phys.} \textbf{\bibinfo{volume}{88}},
  \bibinfo{pages}{6536} (\bibinfo{year}{2000}).

\end{thebibliography}

\end{document}